\documentclass{jfm}
\usepackage{graphicx}
\usepackage{epstopdf, epsfig}
\usepackage{enumitem}
\usepackage{amsmath}
\usepackage{tikz}
\usetikzlibrary{shapes}
\usepackage{scalerel}
\usepackage{stackengine,wasysym}
\usepackage{xcolor}
\usepackage{array}

\shorttitle{Recursive NN-based SGS model for LES}
\shortauthor{Chonghyuk Cho, Jonghwan Park and Haecheon Choi}

\title{A recursive neural-network-based subgrid-scale model for large eddy simulation: application to homogeneous isotropic turbulence}

\author{Chonghyuk Cho\aff{1},
 Jonghwan Park\aff{1}
 \and Haecheon Choi\aff{1,2}
\corresp{\email{choi@snu.ac.kr.}}} 

\affiliation{\aff{1}Department of Mechanical Engineering, Seoul National University, Seoul 08826, Republic of Korea \aff{2}Institute of Advanced Machines and Design, Seoul National University, Seoul 08826, Republic of Korea}
 
\begin{document}

\newcommand{\emptyline}{\raisebox{2pt}{\tikz{\draw[-,black,solid,line width = 0.5pt](0mm,0) -- (4.mm,0)}}}
\newcommand{\blackline}{\raisebox{2pt}{\tikz{\draw[-,black,solid,line width = 1.0pt](0mm,0) -- (8.mm,0)}}}
\newcommand{\dottedline}{\raisebox{2pt}{\tikz{\draw[-,black,dotted,line width = 1.0pt](0mm,0) -- (8.mm,0)}}}
\newcommand{\dashedline}{\raisebox{2pt}{\tikz{\draw[-,black,dashed,line width = 1.0pt](0mm,0) -- (8.mm,0)}}}
\newcommand{\blueline}{\raisebox{2pt}{\tikz{\draw[-,blue,solid,line width = 1.0pt](0mm,0) -- (8.mm,0)}}}
\newcommand{\bluedashedline}{\raisebox{2pt}{\tikz{\draw[-,blue,dashed,line width = 1.0pt](0mm,0) -- (8.mm,0)}}}
\newcommand{\bluedottedline}{\raisebox{2pt}{\tikz{\draw[-,blue,dotted,line width = 1.0pt](0mm,0) -- (8.mm,0)}}}
\newcommand{\redline}{\raisebox{2pt}{\tikz{\draw[-,red,solid,line width = 1.0pt](0mm,0) -- (8.mm,0)}}}
\newcommand{\reddashedline}{\raisebox{2pt}{\tikz{\draw[-,red,dashed,line width = 1.0pt](0mm,0) -- (8.mm,0)}}}
\newcommand{\reddottedline}{\raisebox{2pt}{\tikz{\draw[-,red,dotted,line width = 1.0pt](0mm,0) -- (8.mm,0)}}}
\newcommand{\greenline}{\raisebox{2pt}{\tikz{\draw[-,green,solid,line width = 1.0pt](0mm,0) -- (8.mm,0)}}}
\newcommand{\blasquare}{\raisebox{0.5pt}{\tikz{\node[draw,scale=0.4,regular polygon, regular polygon sides=4,fill=black!100](){};}}}
\newcommand{\blatriangle}{\raisebox{0.5pt}{\tikz{\node[draw,scale=0.3,regular polygon, regular polygon sides=3,fill=black!100!,rotate=0](){};}}}
\newcommand{\bluecircle}{\raisebox{0.5pt}{\tikz{\node[draw,scale=0.4,circle,color=blue,fill=blue](){};}}}
\newcommand{\redcircle}{\raisebox{0.5pt}{\tikz{\node[draw,scale=0.4,circle,color=red,fill=red](){};}}}
\newcommand{\greencircle}{\raisebox{0.5pt}{\tikz{\node[draw,scale=0.4,circle,color=green,fill=green](){};}}}
\newcommand{\blackcircle}{\raisebox{0.5pt}{\tikz{\node[draw,scale=0.4,circle,color=black,fill=black](){};}}}
\newcommand\reallywidetilde[1]{\ThisStyle{%
 \setbox0=\hbox{$\SavedStyle#1$}%
 \stackengine{-.1\LMpt}{$\SavedStyle#1$}{%
  \stretchto{\scaleto{\SavedStyle\mkern.2mu\AC}{.5150\wd0}}{.6\ht0}%
 }{O}{c}{F}{T}{S}%
}}

\maketitle

\begin{abstract}
We introduce a novel recursive process to a neural-network-based subgrid-scale (NN-based SGS) model for large eddy simulation (LES) of high Reynolds number turbulent flow. This process is designed to allow an SGS model to be applicable to a hierarchy of different grid sizes without requiring an expensive filtered direct numerical simulation (DNS) data: 1) train an NN-based SGS model with filtered DNS data at a low Reynolds number; 2) apply the trained SGS model to LES at a higher Reynolds number; 3) update this SGS model with training data augmented with filtered LES (fLES) data, accommodating coarser filter size; 4) apply the updated NN to LES at a further higher Reynolds number; 5) go back to 3) until a target (very coarse) filter size divided by the Kolmogorov length scale is reached. We also construct an NN-based SGS model using a dual NN architecture whose outputs are the SGS normal stresses for one NN and the SGS shear stresses for the other NN. The input is composed of the velocity gradient tensor and grid size. Furthermore, for the application of an NN-based SGS model trained with one flow to another flow, we modify the NN by eliminating bias and introducing leaky rectified linear unit function as an activation function. The present recursive SGS model is applied to forced homogeneous isotropic turbulence (FHIT), and successfully predicts FHIT at high Reynolds numbers. The present model trained from FHIT is also applied to decaying homogeneous isotropic turbulence, and shows an excellent prediction performance. 
\end{abstract}

\section{Introduction}
\label{sec:1}
Large eddy simulation (LES) is an effective tool for predicting turbulent flow because it does not require significant computational resources towards resolution of smaller eddies. This resource reduction is realized through modeling of eddy motions at scales smaller than the grid size (called subgrid scales). The subgrid-scale (SGS) stress from these SGS motions should be modeled to close the governing equations for LES.

The SGS models have been derived for decades based on turbulence theory and statistical approximation. A subset of these models that favor simplicity and stability are represented as linear eddy-viscosity models. Examples include the Smagorinsky model \citep{Smagorinsky63}, based on the theory that the rate of energy transfer to smaller eddies is counterbalanced by the viscous dissipation within the inertial subrange, WALE model \citep{Nicoud99}, using square of the velocity gradient tensor and analyzing proper near-wall scaling for the eddy viscosity, and Vreman model \citep{Vreman04}, rooted in the principle that the SGS stress should be reduced in near-wall region or vanish in laminar flow. Many authors started by \cite{Germano91} have adopted dynamic approaches to extend eddy-viscosity models to more intricate flows. Scale-similarity model (SSM) \citep{Bardina80} assumed that the interplay between resolved and modeled eddies can be accurately delineated by the difference between the filtered and doubly filtered velocities. Lastly, gradient model (GM) \citep{Clark79}) was derived through an approximation technique that employed the Taylor expansion of box filtering. Nevertheless, these models contain some limitations. The linear eddy-viscosity models, in particular, show very low correlations with true SGS stresses in \textit{a priori} tests, irrelevant to the implementation of a dynamic approach, while they work adequately in actual simulations \citep{Liu94, Salvetti95, Park05}. In contrast, both SSM and GM exhibit improved results in \textit{a priori} tests, but they do not sufficiently dissipate turbulent kinetic energy in actual simulations, thereby leading to numerical instability \citep{Vreman96,Vreman97}.

An alternative approach for the derivation of an SGS model involves a direct employment of direct numerical simulation (DNS) data. For example, the optimal LES \citep{Langford99,Volker02} adopted stochastic estimation techniques \citep{Adrian89,Adrian90} to diminish the discrepancy between the ideal and simulated flow variables. This approach necessitated the use of multipoint correlation data as inputs, which in turn required corresponding DNS data. \citet{Moser09} presented an optimal LES that did not require DNS data, but the SGS model was created based on the assumption of isotropic flow.
Another example is an employment of deep learning techniques with a particular emphasis on artificial neural networks (NNs) \citep{Sarghini03,Gamahara17,Maulik18}. Creating an SGS model through NNs requires the accumulation of extensive data through filtering DNS data. This approach is based on an assumption that there may be a complex but well-defined relationship between the SGS stress and resolved flow variables. The pioneering application of NNs in the calculation of SGS stress, to the best of our knowledge, was implemented by \citet{Sarghini03}. They employed an NN to find an optimal turbulent viscosity coefficient for a hybrid model (combined Smagorinsky and similarity models) applied to turbulent channel flow with nine velocity gradients and six resolved Reynolds stresses as inputs, and demonstrated its ability to accurately approximate the coefficient. 

With the rapid advancement and growing recognition of deep learning \citep{Krizhevsky12, Silver16}, more expansive techniques have been employed to develop SGS models for diverse applications. NN-based SGS models may be classified into three categories: closed-form, reassembling, and direct NN-based SGS models, respectively. 
The closed-form NN-based SGS models, analogous to the study conducted by \citet{Sarghini03}, postulate that SGS models may be derived in a closed form from physical, empirical, or analytical tools. The coefficients of these models are adjustable, and are computed using NNs with resolved flow variables as inputs. \citet{Wollblad08} identified the coefficients pertinent to proper orthogonal decomposition for turbulent channel flow through an NN. \citet{Beck19} and \citet{Pawar20} suggested that calculating eddy viscosity could be employed as a surrogate to the direct computation of SGS stress and contribute to enhanced stability. Moreover, \citet{Xie19c}, \citet{Xie20d} and \citet{Wang21} suggested that the SGS stress can be evaluated through a polynomial function including the strain rate, rotation rate, velocity gradient and grid size, with NNs utilized to determine the coefficients of the polynomial. \citet{Yu22} and \citet{Liu23} obtained the coefficients of Smagorinsky and helicity SGS models using NN, respectively.

Reassembling NN-based SGS models extract unfiltered variables from filtered ones, and subsequently utilize these extracted variables for the computation of SGS stress. In this approach, NNs play a prominent role in the unfiltering or filtering process. \citet{Maulik18} engaged NNs in the training for unfiltering \citep{Maulik17} and filtering for two-dimensional decaying homogeneous isotrpoic turbulence (DHIT) in LES. The first NN collected filtered vorticity and streamfunction data across multiple grids and gauged the values of unfiltered vorticity and streamfunction at a single grid. The second NN was subsequently deployed to approximate the filtered nonlinear term, a crucial element for the determination of the SGS stress. Aligning with this methodology, \citet{Yuan20} employed an NN for the estimation of SGS stresses in three-dimensional forced homogeneous isotropic turbulence (FHIT), specifically utilizing the NN to unfilter the velocities.

Direct NN-based SGS models derive the SGS stress or force directly from resolved flow variables through the application of NNs. \citet{Gamahara17} employed NNs to compute the SGS stress as immediate outputs for turbulent channel flow from four input groups such as the strain rate, rotation rate, wall distance, and velocity gradient. Previous direct NN-based SGS models have used various techniques and methods to enhance model performance and stability. These include an \textit{ad hoc} method such as wall-damping function or clipping \citep{Gamahara17, Maulik19, Zhou19}, derivation of inputs from multiple grids for single or multiple outputs \citep{Beck19,Maulik19,Xie19a,Zhou19,Pawar20,Sirignano20,Xie20c,Xie20b,Xie20a,MacArt21,Stoffer21,Cheng22,Guan22,Liu22,Guan23}, incorporation of second derivatives of velocities \citep{Wang18,Xie19a,Pawar20,Sirignano20,Xie20c,MacArt21}, and consideration of wall distance or filter size as inputs \citep{Gamahara17,Zhou19,Abekawa23}. Others \citep{Sirignano20,MacArt21,Guan23,Sirignano23} defined the training error (or objective function) using flow variables other than the SGS stresses. 
\citet{Sirignano20}, \citet{MacArt21} and \citet{Sirignano23} trained NNs with adjoint-based, PDE (partial differential equation)-constrained optimization methods.
Other improvements involve retraining NNs through transfer learning for higher Reynolds number flows \citep{Subel21,Guan22}, and training NNs in complex flows \citep{MacArt21,Sirignano23,Kim2023b}. On the other hand, other types of techniques to improve the performance of NN-based SGS models have also been explored, such as the utilization of convolutional neural network \citep{Beck19,Pawar20,Guan22,Liu22,Guan23}, reinforcement learning \citep{Novati21,Kim22,Kurz23}, and graph neural network \citep{Abekawa23}.

\citet{Park21} developed a direct NN-based SGS model for turbulent channel flow whose input and output were the velocity gradient or strain rate and the SGS stress, respectively. They performed a comprehensive analysis of the salient characteristics inherent to direct NN-based SGS models, and demonstrated from \textit{a posteriori test} without introducing \textit{ad hoc} techniques such as wall-damping function or clipping that SGS models relying on a single grid input can compute turbulence statistics with a reasonably decent level of accuracy. In addition, two specific results from this study are noteworthy: first, when the LES grid size falls within the range of filter sizes for training NN, the SGS model successfully predicts the flow; second, the SGS model is also successful in predicting the flow at higher Reynolds number when the grid size in wall units is same as that trained. Hence, these observations indicate that the applicability of direct NN-based SGS models depends on the dimensionless LES grid size relative to the trained grid (or filter) size. For the development of general NN-based SGS model for complex flow, it is essential that it should function properly over a wide range of non-dimensional filter or grid sizes.

Despite their promising characteristics, previous direct NN-based SGS models have confronted substantial challenges when aimed for application to untrained flows. These challenges originate from the difficulty in developing universal non-dimensional input and output variables for various flows and the nonlinear property (and thus occurrence of potentially large errors in extrapolation) of NN. To achieve regularized input and output values, various techniques have been explored, including min-max method \citep{Xie19a,Pawar20,Wang21}, Gaussian normalization method \citep{Stoffer21,Cheng22,Liu22}, normalization with root-mean-square (rms) values \citep{Xie20a,Xie20b,Xie20c,Guan22,Guan23}, non-dimensionalization in wall units \citep{Park21,Kim22}, and prescribed velocity and length scales \citep{MacArt21}. However, each method has its own limitations: e.g., it requires an equilibrium state or homogeneous direction(s), does not show generalizability when switched to other flows, or proves to be infeasible as it requires DNS results. Consequently, a plausible approach toward general direct NN-based SGS models would be to employ only local variables for non-dimensionalization, like those described in \citet{Prakash22} and \citet{Abekawa23}. 
On the other hand, two strategies may be proposed for the issue of extrapolation errors of NN. The first strategy is to discover normalized flow variables and SGS stresses that are bounded within a certain range for various flows. The second strategy is to incorporate all accessible data to cover an exhaustive range of flow fields. The former presents considerable challenges, because it requires a formidable task of pinpointing normalized flow variables that remain invariant with respect to factors like the Reynolds number, filter size, and flow topology. The latter may demand DNS data of various flows at fairly high Reynolds numbers and thus imposes a huge computational burden.

Therefore, the objective of the present study is to devise a new NN-based SGS model designed to overcome the limitations of existing direct NN-based SGS models. We modify the structure of an NN and formulate a recursive process to generate an SGS model valid for a wide range of grid sizes. Here, an NN is trained using fDNS data at low Reynolds number and then updated by data at higher Reynolds number collected through filtered LES. To validate the performance of the present recursive NN-based SGS model, several SGS models are employed to simulate FHIT at various Reynolds numbers. In addition, two DHIT cases are simulated using the NN trained with FHIT. Section \ref{sec:2} provides the LES framework, NN-based SGS models under consideration, FHIT and filtering method employed. Section \ref{sec:3} describes the recursive algorithm for constructing recursive NN-based SGS models, and provides the results of \textit{a priori} and \textit{a posteriori} tests. In section \ref{sec:4}, the present SGS models are applied to FHIT at high Reynolds numbers and DHIT, respectively, and the results are discussed, followed by conclusions in section \ref{sec:5}.

\section{Numerical details}\label{sec:rules_submission}
\label{sec:2}

\subsection{Large eddy simulation}
In LES, the flow variables are filtered using the following operation:
\begin{equation}
\bar{\phi}(\boldsymbol{x},t)=\int {\bar{G}}(\boldsymbol{r},\boldsymbol{x})\phi(\boldsymbol{x-r},t)d\boldsymbol{r},
\label{eq:one}
\end{equation}
\begin{equation}
\int {\bar{G}}(\boldsymbol{r},\boldsymbol{x})d\boldsymbol{r}=1,
\label{eq:two}
\end{equation}
where $\bar \phi$ is a filtered flow variable, $\bar{G}$ is a filter function, $\boldsymbol{x}$ and $\boldsymbol{r}$ denote position vectors, and $t$ is time. The spatially filtered continuity and Navier-Stokes equations for incompressible flows are
\begin{equation}
\frac{\partial {\bar u}_i}{\partial x_i}=0,
\label{eq:three}
\end{equation}
\begin{equation}
\frac{\partial {\bar u}_i}{\partial t}+\frac{\partial {\bar u}_i {\bar u}_j}{\partial x_j}=-\frac{1}{\rho}\frac{\partial {\bar p^{*}}}{\partial x_i}+\nu  \frac{\partial^2 {\bar u}_i}{\partial {x}_j \partial {x}_j}-\frac{\partial {\tau}^r_{ij}}{\partial {x}_j},
\label{eq:four}
\end{equation}
where $x_i$'s are the coordinates ($x,y,z$), $u_i$'s are the corresponding velocities ($u,v,w$), and $\rho$ and $\nu$ are the fluid density and kinetic viscosity, respectively.
The effect of the SGS eddy motions is modeled as anisotropic part of SGS stress, ${\tau}^r_{ij}$:
\begin{equation}
{\tau}_{ij} = \overline{{ u}_i{ u}_j}-{\bar u}_i{\bar u}_j,
\label{eq:five}
\end{equation}
\begin{equation}
{\tau}^r_{ij} = {\tau}_{ij}-\frac{1}{3}{\tau}_{kk}\delta_{ij},
\label{eq:six}
\end{equation}
where $\delta_{ij}$ is the Kronecker delta. The filtered pressure $\bar p^{*}$ includes the isotropic components of SGS stress:
\begin{equation}
\bar p^{*} = \bar p + \frac{1}{3} \rho \tau_{kk}.
\label{eq:seven}
\end{equation}
In case of HIT, the filtered continuity and Navier-Stokes equations can be transformed in a spectral space. The pseudo-spectral method is utilized for spatial discretization, and the zero-padding method, augmented with the 3/2 rule, is applied to control aliasing errors. For temporal integration, a third-order Runge-Kutta method is used for the convection term, and the second-order Crank-Nicolson method is applied to the diffusion term.

To evaluate the performance of an NN-based SGS model relative to those of traditional SGS models, LESs with the constant Smagorinsky model (CSM, \cite{Smagorinsky63}), dynamic Smagorinsky model (DSM, \cite{Germano91, Lilly92}), and gradient model (GM, \cite{Clark79}) are carried out. The Smagorinsky models determine the anisotropic component of SGS stress from
\begin{equation}
{\tau}^r_{ij} = -2(C_s \bar \Delta)^2 |\bar S| \bar S_{ij},
\label{eq:eight}
\end{equation}
where $C_s$ is the Smagorinsky model coefficient, $\bar \Delta$ is the grid size, $\bar S_{ij}=(1/2)(\partial \bar u_i/\partial {x_j}+\partial \bar u_j/\partial {x_i})$, and $|\bar S|=\sqrt{2\bar S_{ij} \bar S_{ij}}$. For CSM, $C_{s}$ is a fixed value of 0.17. For DSM, the Smagorinsky model coefficient is obtained as 
\begin{equation}
(C_s \bar \Delta)^2 = \max \left[-\frac{1}{2}\langle L_{ij}M_{ij} \rangle_h/ \langle M_{ij}M_{ij} \rangle_h, 0\right],
\label{eq:nine}
\end{equation}
where
\begin{equation}
L_{ij}=\widetilde{\bar u_i \bar u_j} - {\widetilde{\bar u}_i} {\widetilde{\bar u}_j},
\label{eq:ten}
\end{equation}
\begin{equation}
M_{ij}=\left(\frac{\widetilde {\Delta}}{\bar \Delta}\right)^2 |\widetilde{\bar S}| \widetilde{\bar S}_{ij} - \widetilde{|\bar S|\bar S_{ij}},
\label{eq:eleven}
\end{equation}
$\overline{(\cdot)}$ and $\widetilde{(\cdot)}$ correspond to the grid- and test-filtering operations, respectively, and $\widetilde{\Delta} = 2\bar \Delta$. The notation $\langle {\cdot} \rangle_h$ denotes an instantaneous averaging over homogeneous directions.
The SGS stress in GM is derived through the application of Taylor expansions of the filtered velocity \citep{Clark79, Vreman96} as
\begin{equation}
\tau_{ij}=\frac{1}{12}\sum\limits_{k} {\bar \Delta_k^2} \frac{\partial \bar u_i}{\partial{x}_k} \frac{\partial \bar u_j}{\partial{x}_k},
\label{eq:twelve}
\end{equation}
and $\tau_{ij}^r$ is computed from (\ref{eq:six}), where $\bar \Delta_k$ is the grid size in $x_k$ direction.

\subsection{NN-based SGS model}
So far, most NN-based SGS models have optimized the weights and biases of the NNs, despite utilizing different deep learning techniques (see \S \ref{sec:1}). For example, \citet{Park21} implemented an NN consisting of two hidden layers and 128 neurons per hidden layer to compute six components of the SGS stress: 
\begin{equation}
\left. \begin{array}{ll}  
h_i^{(1)}= \bar q_i & (i=1,2,\cdots,N_q); \\
h_j^{(2)}=\max \left[0,\gamma_{j}^{(2)} \left( \sum\limits_{i=1}^{N_q}W^{(1)(2)}_{ij}h^{(1)}_i + b_{j}^{(2)} - \mu_{j}^{(2)} \right) / \sigma_{j}^{(2)} + \beta_{j}^{(2)} \right] & (j=1,2,\cdots ,128); \\ \\
h_k^{(3)}=\max \left[0,\gamma_{k}^{(3)} \left( \sum\limits_{j=1}^{128}W^{(2)(3)}_{jk}h^{(2)}_j + b_{k}^{(3)} - \mu_{k}^{(3)} \right) / \sigma_{k}^{(3)} + \beta_{k}^{(3)}\right] & (k=1,2,\cdots ,128); \\ \\
h_l^{(4)}=s_l=\sum\limits_{k=1}^{128}W^{(3)(4)}_{kl}h^{(3)}_k + b_{l}^{(4)} & (l=1,2,\cdots ,6).\\
\end{array}\right\}
\label{eq:thirteen}
\end{equation}
Here, $\bar{\boldsymbol{q}}$ is the grid-filtered input, $N_q$ is the number of the input components, $\boldsymbol{W}^{(m)(m+1)}$ is the weight matrix between $m$th and $(m+1)$th layers, $\boldsymbol{b}^{(m)}$ is the bias vector of the $m$th layer, $\boldsymbol{s}$ is the output (six components of the SGS stress), and $\boldsymbol{\gamma}^{(m)}$, $\boldsymbol{\mu}^{(m)}$, $\boldsymbol{\sigma}^{(m)}$ and $\boldsymbol{\beta}^{(m)}$ are the parameters for batch normalization \citep{Ioffe15}.
During an NN training process, the parameters $\boldsymbol{W}^{(m)(m+1)}$ and $\boldsymbol{b}^{(m+1)}$ are optimized to minimize the training error. 

\begin{figure}
 \centerline{\includegraphics[width=0.9\textwidth]{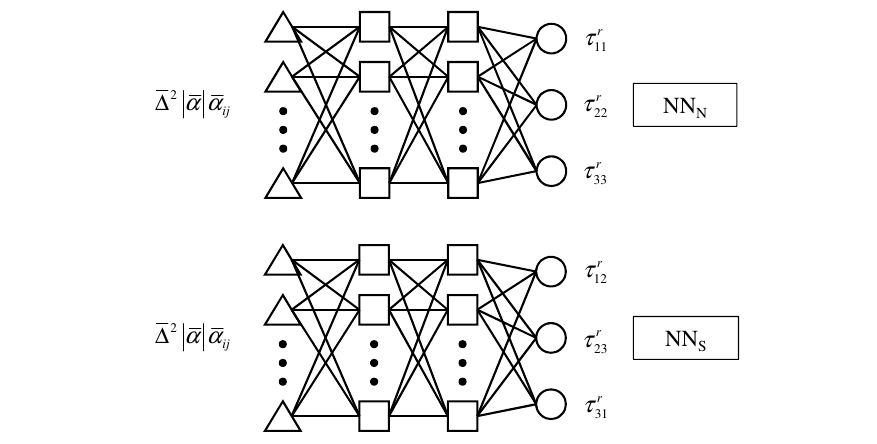}}
 \caption{Schematic diagram of the present NN-based SGS model that consists of a dual NN: NN$_{\rm N}$ and NN$_{\rm S}$ predict the SGS normal and shear stresses, respectively.}
\label{fig:NN}
\end{figure}

In the present study, we construct an NN-based SGS model using a dual NN architecture (figure \ref{fig:NN}), where the output of one NN (NN$_{\rm N}$) is the SGS normal stresses $(\tau_{11}^r,\tau_{22}^r,\tau_{33}^r)$, and that of the other (NN$_{\rm S}$) is the SGS shear stresses $(\tau_{12}^r,\tau_{13}^r,\tau_{23}^r)$. The reason of using two NNs is that the ranges of the normal and shear SGS stresses are quite different from each other and thus separate treatments of these SGS stresses increase the prediction capability of the present NN-based SGS model. 
Each of the present NNs consists of two hidden layers and 64 neurons per hidden layer, and the output of the $m$-th layer, $\boldsymbol{h}^{(m)}$, is computed as
\begin{equation}
\left.\begin{array}{ll}
h_i^{(1)}= \bar q_i & (i=1,2,\cdots,9); \\
h_j^{(2)}=\max \left[0.02r_j^{(2)},r_j^{(2)}\right], \quad r_j^{(2)}=\sum\limits_{i=1}^{9}W^{(1)(2)}_{ij}h^{(1)}_i  & (j=1,2,\cdots ,64); \\ \\
h_k^{(3)}=\max \left[0.02r_k^{(3)},r_k^{(3)} \right], \quad r_k^{(3)}=\sum\limits_{j=1}^{64}W^{(2)(3)}_{jk}h^{(2)}_j & (k=1,2,\cdots ,64); \\ \\
h_l^{(4)}=s_l=\sum\limits_{k=1}^{64}W^{(3)(4)}_{kl}h^{(3)}_k & (l=1,2,3).
\end{array} \right\}
\label{eq:fourteen}
\end{equation}
Note that we remove the bias $\boldsymbol{b}$ and parameters for batch normalization ($\boldsymbol{\gamma}$, $\boldsymbol{\mu}$, $\boldsymbol{\sigma}$ and $\boldsymbol{\beta}$). Also, we use the leaky rectified linear unit function (leaky ReLU) as an activation function: $f(x) = \max[ax, x]$, which was proposed by \citet{Maas13} to overcome vanishing gradient problem \citep{Bengio94}. The negative slope is determined to be $0.01 \leq a \leq 0.2$ \citep{Xu15}, and we choose $a=0.02$. 
Without the bias and batch-normalization parameters and with the use of leaky ReLU, the present NN complies with the following conditions:
\begin{equation}
\boldsymbol{{\mathrm{NN}}} (c\boldsymbol{A}) = c \boldsymbol{{\mathrm{NN}}} (\boldsymbol{A}) \quad \mathrm{only \ if } \ c \geq 0,
 \label{eq:sixteen}
\end{equation}
\begin{equation}
\boldsymbol{{\mathrm{NN}}} (\boldsymbol{A}+\boldsymbol{B}) \neq \boldsymbol{{\mathrm{NN}}} (\boldsymbol{A}) + \boldsymbol{{\mathrm{NN}}} (\boldsymbol{B}),
 \label{eq:seventeen}
\end{equation}
where 
$c$ is a positive scalar, and $\boldsymbol{A}$ and $\boldsymbol{B}$ are arbitrary tensors. 
Thus, the present NN still retains its nonlinearity, making it suitable for nonlinear regression between the SGS stresses and local flow variables. 
Equation (\ref{eq:sixteen}) allows us to apply an NN trained from one flow to another flow (see \S \ref{sec:3_1}).

During the training process, the weight parameters $\boldsymbol{W}^{(m)(m+1)}$ are optimized to minimize the mean square error given by
\begin{equation}
L=\frac{1}{3}\frac{1}{N_{b}}\sum\limits_{l=1}^{3} \sum\limits_{n=1}^{N_b}(s_{l,n}^{\mathrm{fDNS}}-s_{l,n})^2,
\label{eq:fifteen}
\end{equation}
where $N_{b}$ denotes the minibatch size of 256, and $s_{l,n}^{\mathrm{fDNS}}$ corresponds to the anisotropic components $\tau^r_{ij}$ obtained from fDNS. 

We apply Adam algorithm (a type of gradient descent, \cite{Kingma14}) and learning rate annealing method \citep{Simonyan14,He16}
to optimize the weights and enhance the training speed, respectively. Early stopping \citep{Goodfellow16} is employed in the training process to avoid overfitting. The learning rate is initially 0.025, and is subsequently reduced by a factor of 10 if no improvement in the training error is observed over five epochs. If the learning rate is decreased three times but there is no reduction in the training error over five epochs, training is stopped. The entire process to train and execute the present NN is carried out with the PyTorch open-source library in the Python programming environment.

We develop an NN-based SGS model (velocity gradient model, called VGM hereafter) whose input and output are $\bar \Delta^2|\bar \alpha|\bar \alpha_{ij}$ and $\tau^r_{ij}$, respectively, where $\bar \Delta = (\bar \Delta_x \bar \Delta_y \bar \Delta_z)^{1/3}$, $\bar \alpha_{ij}=\partial \bar u_i / \partial x_j$, and $|\bar \alpha|=\sqrt{\bar \alpha_{ij}\bar \alpha_{ij}}$. Note that the dimensions of input and output are matched to be the same. The velocity gradient tensor as an input has been used by many previous studies \citep{Gamahara17,Wang18,Zhou19,Xie19a,Pawar20,Prat20,Xie20a,Xie20b,Xie20c,Park21,Kim22,Abekawa23}, and the same form of input was considered by \citet{Jamaat22} for one-dimensional Burgers turbulence. 
The VGM obtains inputs from each grid point and compute corresponding SGS stresses.

\subsection{Forced homogeneous isotropic turbulence (FHIT) and filtering method}
FHIT involves an additional forcing term in the Navier-Stokes equations at low wavenumbers \citep{Ghosal95,Rosales05,Park06}: 
\begin{equation}
  f_{i} = \epsilon_{t}~ \frac{u_{i}}{\sum_{0<\vert \mathbf{k} \vert <2} |\hat{u}_{i}(\mathbf{k})|^{2}},
 \label{eq:lowfFHIT1}
\end{equation}
 where $f_{i}$ is the forcing term, $\hat u_i$ is the Fourier coefficient of $u_i$, $\epsilon_{t}$ is the prescribed mean total dissipation rate determining the turbulent energy injection rate, $\vert \mathbf{k} \vert =\sqrt{k_x^2 + k_y^2 + k_z^2}$, and $k_i$ is the wavenumber in $i$ direction. Note that $\epsilon_t$ encompasses both the resolved and modeled dissipation.

\begin{table}
 \begin{center}
 \begin{tabular}{lrrr}
  \quad Case & \quad $N_{\text{DNS}}$ \quad & $Re_L$ \quad & \quad $Re_\lambda$ \quad \\
  DNS128 & 128 & 149.09 & \qquad 93.03\\
  DNS256 & 256 & 375.69 & 154.17\\
  DNS512 & 512 & 946.67 & -\\
  DNS1024 & 1024 & 2385.46 & -\\
  DNS2048 & 2048 & 6010.98 & -\\
  DNS4096 & 4096 & 15146.71 & -\\
  DNS8192 & 8192 & 38167.31 & -\\
  DNS16384 & 16384 & 96175.60 & -\\
  DNS32768 & \qquad 32768 & \qquad 242347.33 & -\\
 \end{tabular}
 \caption{Cases of DNS at various Reynolds numbers for FHIT. Here, $Re_L = (N_{\rm DNS}/3)^{4/3}$ and $\Delta_{\rm DNS}/\eta = 2\pi / 3 ~(N_{\rm DNS} \Delta_{\rm DNS} = 2\pi L)$. DNSs are performed for DNS128 and DNS256, and the corresponding $Re_\lambda$'s are given in this table.}
 \label{tab:table2}
 \end{center}
\end{table}

In FHIT, two distinct Reynolds numbers, $Re_{\lambda}=u_{\text{rms}}{\lambda}/\nu$ and $Re_L = UL/\nu$, can be defined. Here, $u_{\text{rms}}$ is the rms velocity fluctuations, $\lambda (= \sqrt{15 u_{\text{rms}}^2 \nu / \epsilon_t})$ is the Taylor microscale, $L$ is associated with the computational domain size of $2\pi L \times 2\pi L \times 2\pi L$, and $\epsilon_t$ is normalized to unity such that $\epsilon_t L/U^{3} = 1$. 
The Taylor microscale Reynolds number $Re_{\lambda}$ has been widely used in turbulence studies. Nevertheless, the extraction of $u_{rms}$ requires intricate experimental data or DNS.
In contrast, $Re_L$ does not require such preliminary outcomes, and can be deduced from $\epsilon_t L/U^{3} = 1$, $\eta k_{\text{max,DNS}}$ and $N_{\text{DNS}}$, where $\eta$ is the Kolmogorov length scale, and $k_{\text{max,DNS}}$ and $N_{\text{DNS}}$ are the largest wavenumber and number of grid points in each direction in DNS, respectively. 
According to \citet{Pope00}, $\eta k_{\text{max,DNS}}=3/2$ or $\Delta_{\text{DNS}}/\eta = 2\pi / 3$ is set to achieve sufficiently high resolution in DNS, where $\Delta_{\text{DNS}}$ is the grid size in DNS. Then, $L = \Delta_{\text{DNS}} N_{\text{DNS}} / (2\pi) = \eta N_{\text{DNS}}/3$. With these relations together with $\eta = (\nu^3 / \epsilon_t )^{1/4}$, one can obtain $Re_L = (N_{\text{DNS}} / 3)^{4/3}$. Table \ref{tab:table2} lists the cases of DNS at various Reynolds numbers by changing the number of grid points for FHIT, where each case is named as DNS$N_{\text{DNS}}$.
Note that the cases listed in table \ref{tab:table2} do not necessarily require actual DNS except for the case of DNS128 (the recursive procedure introduced in the below requires DNS only at a low Reynolds number), and LESs are conducted for higher Reynolds number cases through the recursive procedure.

\begin{table}
 \begin{center}
 \begin{tabular}{lcrr}
 \qquad \qquad Case & \quad  ~$N_c$ \quad & $Re_L$ \quad & $\bar \Delta / \eta$ \\
  fDNS32/128 or LES32/128 & \quad 32 & 149.09 & 8.38 \\
  fDNS64/128 & \quad 64 & 149.09 & 4.19  \\
  fDNS32/256 or LES32/256 & \quad 32 & 375.69 & 16.76  \\
  fDNS64/256 or LES64/256 & \quad 64 & 375.69 & 8.38  \\
  fDNS128/256 & \quad 128 & 375.69 & 4.19  \\
  \qquad  \qquad \qquad ~~~LES64/512 & \quad 64 & 946.67 & 16.76  \\
  \qquad  \qquad \qquad ~~~LES64/1024 & \quad 64 & \qquad 2385.46 & \qquad 33.51  \\
 \end{tabular}
 \caption{Cases of fDNS and LES at various Reynolds numbers for FHIT. Here, $N_c$ is the number of grid points in each direction for fDNS or LES, and $\bar \Delta$ denotes the filter size for fDNS ($\bar \Delta_{\rm fDNS}$) or the grid size for LES ($\bar \Delta_{\rm LES}$). The name of each case indicates fDNS$N_c / N_{\rm DNS}$ or LES$N_c / N_{\rm DNS}$. Note that $\Delta_{\rm DNS}/\eta = 2\pi / 3$.}
 \label{tab:table3}
 \end{center}
\end{table}

Table \ref{tab:table3} shows the cases considered for fDNS and LES of FHIT.
The fDNS data are obtained by applying a transfer function $\hat{G}(\mathbf{k})$ to the Fourier-transformed DNS data $\hat u_i(\mathbf{k})$: 
\begin{equation}
\hat{\bar u}_{i}(\mathbf{k}) = \hat{u}_{i}(\mathbf{k}) \hat{\bar{G}}(\mathbf{k}),
 \label{eq:twentytwo}
\end{equation}
\begin{equation}
\hat{\bar{G}}(\mathbf{k}) = \int \exp(i \mathbf{k} \cdot \mathbf{r}) \bar{G} (\mathbf{r})d\mathbf{r}.
 \label{eq:twentyone}
\end{equation}
The fDNS data can be obtained by applying a spectral cut-off filter or Gaussian filter in the spectral space. However, these filterings have limitations when the filtered variables are compared with LES results.
The spectral cut-off filtering removes the Fourier coefficients $\hat u_i$ at $k > k_c$, and allows filtered variables to be placed on coarser grids of LES, where $k_c (=\pi / \bar \Delta)$ is the cut-off wavenumber. However, this filtering fails to satisfy the realizability condition due to negative weights in the physical space \citep{Vreman94}. Additionally, the spectral cut-off filtering may exhibit non-local oscillations \citep{Meneveau00, Pope00}. 
On the other hand, the Gaussian filtering is free from these drawbacks, and requires filtering even when the wavenumber exceeds the LES limit (i.e., $k > \pi/\bar \Delta_{\rm LES}$). Although the filter size may coincide with the LES grid size, the Gaussian filtered data are allocated at DNS grids, not at LES grids.

\begin{table}
 \begin{center}
  \begin{tabular}{lcc}
   Filter & $\hat{\bar{G}}(\mathbf{k})$ & Largest wavenumber of filtered variables \\
   \hline
   Spectral cut-off & \quad $H(k_{c} - k)$ \quad & $k_{c}$ \\
   Gaussian & \quad $\exp\left(-k^2 \bar \Delta^2 / 24 \right)$ \quad & $k_{\rm DNS} (=\pi / \Delta_{\rm DNS})$ \\
   Cut-Gaussian & \quad $\exp\left(-k^2 \bar \Delta^2 / 24 \right)$ \quad  & $k_{c}$ \\
  \end{tabular}
  \caption{Comparison of three filters. Here, $H$ is the Heaviside step function.}
  \label{tab:table4}
 \end{center}
\end{table}

\begin{figure}
 \centerline{\includegraphics[width=0.9\textwidth]{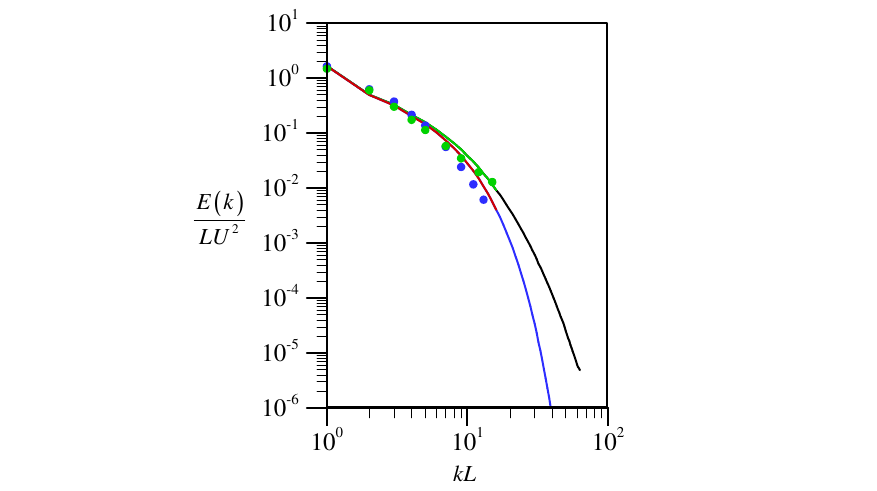}}
 \caption{Three-dimensional energy spectra: \protect\blackline, DNS128 ($Re_L =149.09; Re_\lambda = 93.03$); \protect\redline, fDNS32/128 with cut-Gaussian filtering; \protect\blueline, fDNS32/128 with Gaussian filtering; \protect\greenline, fDNS32/128 with spectral cut-off filtering; \protect\bluecircle, LES32/128 with DSM; \protect\greencircle, LES32/128 with GM.}
\label{fig:filtering}
\end{figure}

In the present study, we use the cut-Gaussian filtering \citep{Zanna20,Guan22,Guan23,Pawar23} which retains the transfer function of Gaussian filtering but truncates the filtered variables at the wavenumbers exceeding the cut-off wavenumber $k_{c}$. Table \ref{tab:table4} shows three different filters in the spectral space, corresponding transfer functions, and largest wavenumbers of filtered variables, respectively.
Figure \ref{fig:filtering} shows the energy spectra from DNS128, fDNS32/128's using cut-Gaussian, Gaussian and spectral cut-off filters, and LES32/128's with DSM and GM, respectively. The energy spectrum with GM is in an excellent agreement with that of fDNS using the spectral cutoff filter. In contrast, the energy spectrum with DSM is closer to that of fDNS using the cut-Gaussian filter. Given the increased adaptability of DSM over GM, we measure the prediction capability of the SGS models based on the cut-Gaussian filtered DNS data.

 \section{A recursive NN-based SGS model}
 \label{sec:3}

\subsection{A recursive algorithm}\label{sec:3_0}
The present study employs a recursive method to construct an NN-based SGS model adapted to various grid sizes normalized by the Kolmogorov length scale from the following steps:
\begin{enumerate}[widest=99, leftmargin=3em, rightmargin=3em, labelwidth=!, labelindent=1em, align=left, itemindent=0pt, itemsep=0pt, parsep=0pt, before={\vspace{1em}}, after={\vspace{1em}}]
\item Obtain training data (fDNS data) by filtering DNS data  at a low Reynolds number ($Re_L$); \label{item:first}
\item Train an NN-based SGS model with the fDNS data; \label{item:second}
\item Apply the NN-based SGS model to LES at a higher Reynolds number, where the ratio of LES grid size to the Kolmogorov length scale ($\bar \Delta_{\rm LES}/\eta$) is equal to that of filter size to the Kolmogorov length scale ($\bar \Delta_{\rm fDNS}/\eta$); \label{item:third}
\item Filter the LES data and include the filtered LES (fLES) data in the training dataset, where the new filter size is twice the LES grid size;  \label{item:fourth}
\item Train the NN-based SGS model using the augmented training data; \label{item:fifth}
\item Apply the updated NN-based SGS model to LES at even higher Reynolds number, where the LES grid size is equal to the filter size defined in step (\ref{item:fourth}); 
\label{item:sixth}
\item Repeat from step (\ref{item:fourth}) to (\ref{item:sixth}) for LES at higher Reynolds numbers.
\label{item:seventh}
\end{enumerate}
\noindent
The steps (\ref{item:first}) and (\ref{item:second}) are similar to those used in constructing previous NN-based SGS models. The present recursive method aims at improving the performance of an NN-based SGS model at a wider range of non-dimensional grid sizes by recursively performing LES and training it using fLES data.

\subsection{NN-based SGS model trained with fDNS data: \textit{a priori} and \textit{a postriori} tests}
\label{sec:3_1}
First, fDNS data is used to train an NN. The Reynolds number and number of grid points in each direction are $Re_{L} = 149.09$ and $N_{\rm DNS}=128$, respectively (table \ref{tab:table2}). The filter sizes considered are $\bar \Delta_{\rm fDNS}/\eta = 8.38$ and $4.19$, corresponding to the cases of fDNS32/128 and fDNS64/128, respectively (table \ref{tab:table3}). 
Including two different filter sizes in constructing fDNS data helps to improve the prediction capability of NN-based SGS models (see, for example, \cite{Park21}) by broadening the ranges of the input and output, when they are appropriately normalized. 
In the present study, we train a dual NN (NN$_{\rm N}$ and NN$_{\rm S}$ in figure \ref{fig:NN}) with the training data (input and output) normalized by rms SGS normal ($\tau_{11, rms}^r$) and shear ($\tau_{12, rms}^r$) stress fluctuations, respectively. Note that $\tau_{11, rms}^r (=\tau_{22, rms}^r = \tau_{33, rms}^r)$ and $\tau_{12, rms}^r (=\tau_{13, rms}^r = \tau_{23, rms}^r)$ are obtained from (\ref{eq:five}) and (\ref{eq:six}) with DNS data. 
During actual LES, however, the SGS normal and shear stress fluctuations are \textit{a priori} unknown, and thus providing normalized input data to the NN is not possible. 
Hence, in actual LES, we can only provide the input normalized by the characteristic velocity scale ($\bar{\Delta}^2|\bar{\alpha}|\bar{\alpha}_{ij}/U_{\rm LES}^2$). Thanks to the important property of the present NN, (\ref{eq:sixteen}), we obtain the following relation: (no summation on $i$ and $j$)
\begin{equation}
\frac{\tau^r_{ij}}{U_{\rm LES}^2} = \frac{\tau_{ij, rms}^r}{U^2_{\rm LES}} \frac{\tau^r_{ij}}{\tau_{ij, rms}^r} = \frac{\tau_{ij, rms}^r}{U^2_{\rm LES}} \boldsymbol{\mathrm{NN}} \left(
\frac{\bar \Delta^2|\bar \alpha| \bar \alpha_{ij}}{\tau_{ij, rms}^r} \right) = \boldsymbol{\mathrm{NN}} \left(
\frac{\bar \Delta^2|\bar \alpha| \bar \alpha_{ij}}{U^2_{\rm LES}} \right).
 \label{eq:prove1}
\end{equation}
This relation indicates that, during actual LES, one can provide the input normalized by $U_{\rm LES}$ to the NN trained with the input normalized by the rms SGS stresses.

\begin{figure}
 \centerline{\includegraphics[width=0.9\textwidth]{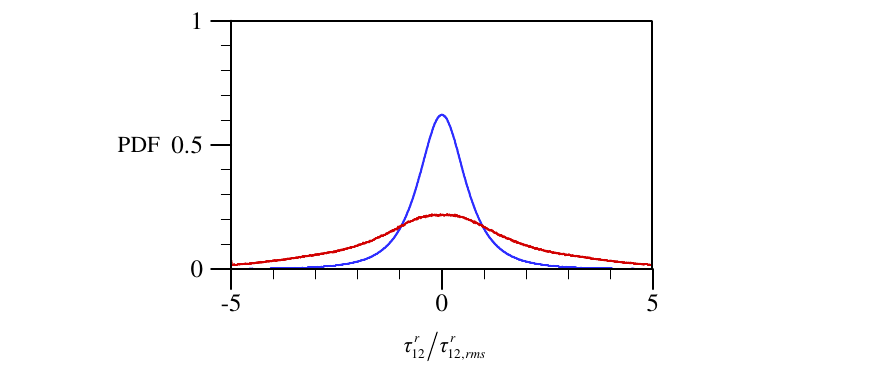}}
 \caption{Probability density function (PDF) of normalized $\tau_{12}^r$ before (\protect\blueline) and after (\protect\redline) undersampling.}
\label{fig:undersamplig}
\end{figure}

On the other hand, the SGS stresses obtained from DNS data have a problem of imbalanced distribution because they tend to be mainly distributed around zero. So, we choose the normalized output and corresponding input through the process known as undersampling \citep{Drummond03,Liu08}. During this process, we do not use all the filtered data as training one, but choose data such that the possibility of choosing data as training one decreases when its magnitude is near to zero: i.e., the probability ($\mathcal{P}$) to choose an SGS shear stress and corresponding inputs as training data is
\begin{equation}
\begin{gathered}
 \left. \begin{aligned}
& \mathcal{P} = \sin^{2}{\theta} &\text{if } \theta < \frac{\pi}{2}\\
& \mathcal{P} = 1 &\text{if } \theta \geq \frac{\pi}{2}
\end{aligned} \right\},\\
 \theta = \frac{\pi}{8} \frac{\sqrt{\left\{(\tau_{12}^{r})^{2}+(\tau_{23}^{r})^{2}+(\tau_{13}^{r})^{2}\right\}/3}}{\tau_{12,rms}^{r}}.
\label{eq:twentythree} 
\end{gathered}
\end{equation}
Figure \ref{fig:undersamplig} illustrates the effect of undersampling on the distribution of training data. This undersampling reduces the probability of zero SGS stresses, and enhances the occurrence of strong SGS stresses, leading to generate high SGS dissipation and thus improving the performance of an NN-based SGS model. The SGS normal stresses are also similarly processed. For each fDNS dataset, about $500,000$ pairs of $\bar \Delta^2 |\bar \alpha| \bar \alpha_{ij}$ and $\tau_{ij}^r$ are used for training.

\begin{table}
  \begin{center}
  \begin{tabular}{clcccc}
  $Re_L$ & SGS model  & $R(\tau_{11}^{r})$ & $R(\tau_{12}^{r})$ & $R(\epsilon_{SGS})$ & $\epsilon_{SGS}$ \\ 
  \hline
 149.09 & fDNS32/128 & \protect\emptyline & \protect\emptyline & \protect\emptyline & 0.430 \\
  & VGM(32+64)/128 \qquad & 0.661 & 0.692 & 0.647 & 0.439 \\
  & CSM32/128 & 0.191 & 0.204 & 0.569 & 1.077 \\
  & DSM32/128 & 0.191 & 0.204 & 0.569 & 0.859 \\
  & GM32/128 & 0.653 & 0.682 & 0.633 & 0.315 \\
  \hline
 375.69 & fDNS64/256 & \protect\emptyline & \protect\emptyline &  \protect\emptyline & 0.439 \\
   &   VGM(32+64)/128 & 0.664 & 0.695 & 0.662 & 0.453 \\
   &   VGM(64+128)/256 \qquad & 0.664 & 0.695 & 0.662 & 0.449 \\
   &   CSM64/256 & 0.186 & 0.201 & 0.578 & 1.118 \\
   &   DSM64/256 & 0.186 & 0.201 & 0.577 & 0.892 \\
   &   GM64/256 & 0.656 & 0.686 & 0.647 & 0.324 \\
  \end{tabular}
 \caption{Statistics from \textit{a priori} tests at $Re_L=149.09$ and 375.69 with $\bar \Delta_{\rm fDNS}/\eta=8.38$: Pearson correlation coefficients $(R)$ of $\tau_{ij}^{r}$ and SGS dissipation $\epsilon_{SGS} (=-\tau_{ij}^{r}\bar S_{ij})$, and magnitude of $\epsilon_{SGS}$.}
  \label{tab:table5}
  \end{center}
\end{table}

\begin{figure}
 \centerline{\includegraphics[width=0.9\textwidth]{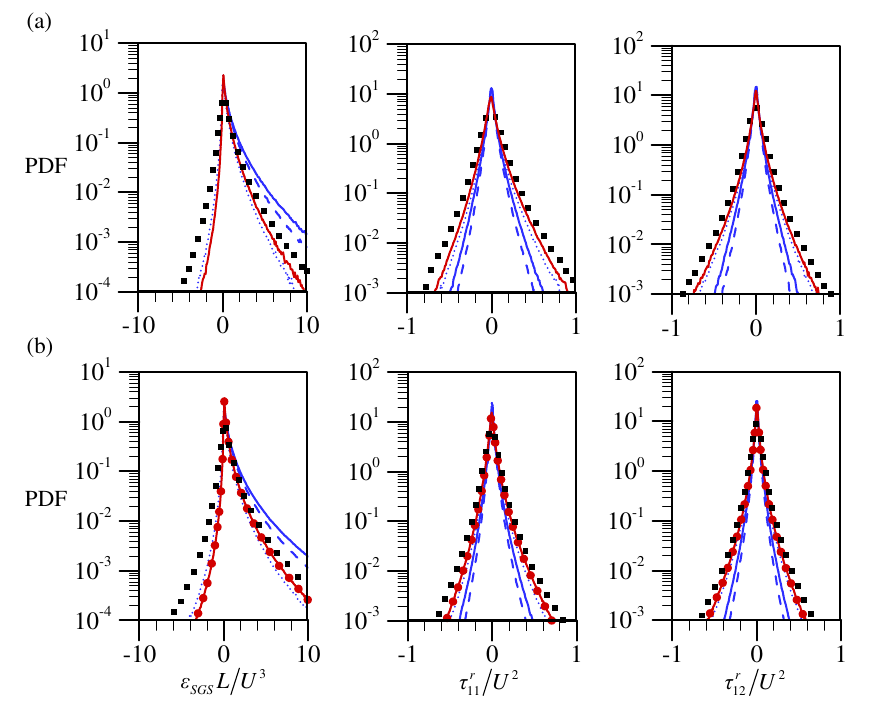}}
 \caption{Probability density functions of the SGS dissipation and SGS normal and shear stresses (\textit{a priori} test with $\bar \Delta_{\rm fDNS}/\eta=8.38$): (a) $Re_L=149.09$; (b) $Re_L=375.69$. In (a), \protect\blasquare, fDNS32/128; \protect\redline, VGM(32+64)/128;  \protect\blueline, CSM32/128; \protect\bluedashedline, DSM32/128; \protect\bluedottedline, GM32/128.  In (b), \protect\blasquare, fDNS64/256; \protect\redcircle, VGM(64+128)/256; \protect\redline, VGM(32+64)/128; \protect\blueline, CSM64/256; \protect\bluedashedline, DSM64/256; \protect\bluedottedline, GM64/256.}
\label{fig:priori1}
\end{figure}

We perform \textit{a priori} tests with the present VGMs and traditional models using fDNS32/128 ($Re_L = 149.09$) and fDNS64/256 ($Re_L = 375.69$), respectively. Table \ref{tab:table5} provides the correlations of the SGS normal and shear stresses and SGS dissipation, and the magnitude of the SGS dissipation from various SGS models, respectively. Here, VGM(32+64)/128 is trained using fDNS32/128 and fDNS64/128 datasets, and VGM(64+128)/256 is done using fDNS64/256 and fDNS128/256 datasets, respectively. For both Reynolds numbers, VGMs provide the highest correlations of the SGS stresses and the magnitudes of the SGS dissipation closest to those of fDNS data. 
It is notable to see that VGM(32+64)/128 trained at $Re_L = 149.09$ successfully predicts the correlations and SGS dissipation even when it is applied to a higher Reynolds number of $Re_L = 375.69$. 

Figures \ref{fig:priori1}(a) and (b) show the PDFs of $\epsilon_{SGS}$, $\tau_{11}^{r}$ and $\tau_{12}^{r}$ from various SGS models for $Re_L = 149.09$ and 375.69, respectively. As shown, the PDFs of $\epsilon_{SGS}$ from VGM and GM agree very well with fDNS data. The PDFs from both CSM and DSM appear only at positive SGS dissipation but do not agree well with fDNS data. Similarly, for SGS stresses, VGM and GM work very well, whereas CSM and DSM are not good. It should be noted that VGM(32+64)/128 trained at $Re_L=149.09$ predicts the PDFs accurately at $Re_L=375.69$  as much as VGM(64+128)/256 does.

\begin{figure}
  \centerline{\includegraphics[width=1.0\textwidth]{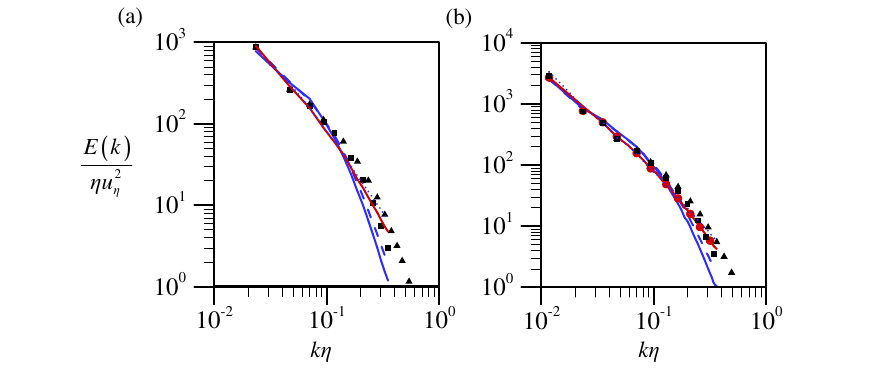}}
  \caption{Three-dimensional energy spectra (\textit{a posteriori} test; LES32/128 and LES64/256 for $Re_L=149.09$ and 375.69, respectively, with $\bar \Delta_{\rm LES}/\eta=8.38$): (a) $Re_L=149.09$; (b) $Re_L=375.69$. In (a), \protect\blatriangle, DNS128; \protect\blasquare, fDNS32/128; \protect\redline, VGM(32+64)/128; \protect\blueline,  CSM32/128; \protect\bluedashedline,  DSM32/128; \protect\bluedottedline, GM32/128. In (b), \protect\blatriangle, DNS256; \protect\blasquare, fDNS64/256; \protect\redcircle, VGM(64+128)/256; \protect\redline, VGM(32+64)/128; \protect\blueline, CSM64/256; \protect\bluedashedline, DSM64/256; \protect\bluedottedline, GM64/256. Here, $u_\eta$ is the Kolmogorov velocity scale.}
\label{fig:Ek1d1}
\end{figure}

\begin{figure}
  \centerline{\includegraphics[width=0.9\textwidth]{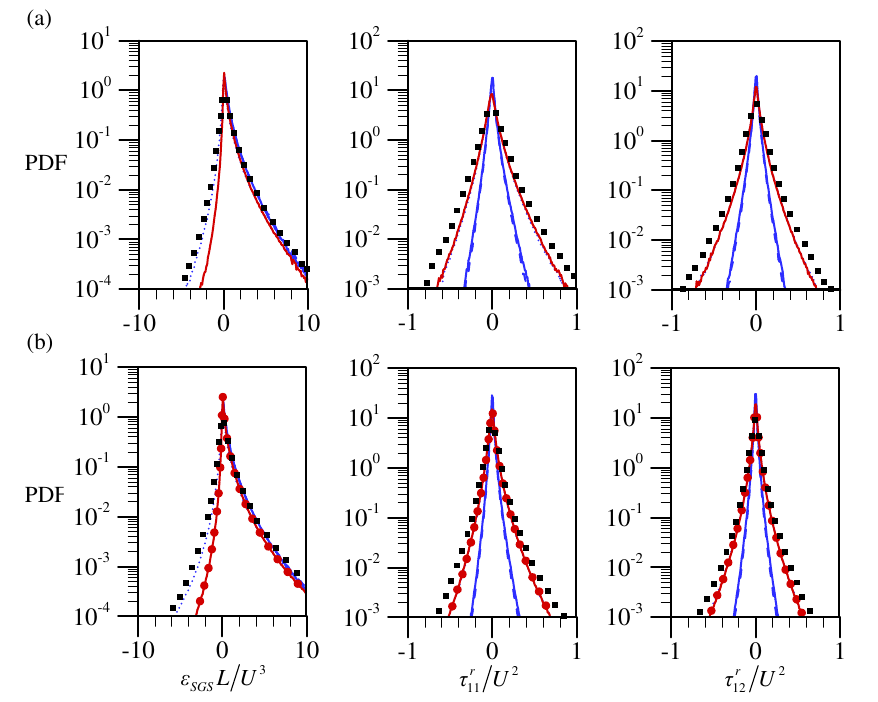}}
  \caption{Probability density functions (\textit{a posteriori} test; LES32/128 and LES64/256 for $Re_L=149.09$ and 375.69, respectively, with $\bar \Delta_{\rm LES}/\eta=8.38$): (a) $Re_L=149.09$; (b) $Re_L=375.69$. In (a), \protect\blasquare, fDNS32/128; \protect\redline, VGM(32+64)/128; \protect\blueline, CSM32/128; \protect\bluedashedline, DSM32/128; \protect\bluedottedline, GM32/128. In (b), \protect\blasquare, fDNS64/256; \protect\redcircle, VGM(64+128)/256; \protect\redline, VGM(32+64)/128; \protect\blueline, CSM64/256; \protect\bluedashedline, DSM64/256; \protect\bluedottedline, GM64/256.}
\label{fig:posteriori1}
\end{figure}

Now, we perform \textit{a posteriori} tests (actual LES) with the present VGMs and traditional models for $Re_L=149.09$ and 375.69, and compare the results with fDNS32/128 and fDNS64/256 data, respectively. Figures \ref{fig:Ek1d1} and \ref{fig:posteriori1} show the three-dimensional energy spectra and PDFs of the SGS dissipation and SGS stresses, respectively. The energy spectra from VGM and GM agree well with fDNS data and better than those of CSM and DSM. For PDFs, GM provides the most accurate predictions and VGM also performs very well, whereas those from CSM and DSM are not good. 
These results indicate that LES with VGM trained at a Reynolds number predicts turbulence statistics very well for a different (higher) Reynolds number flow if $\bar \Delta_{\rm LES}/\eta = \bar \Delta_{\rm fDNS}/\eta$.
A similar conclusion was also made by \citet{Park21}, where the grid size was non-dimensionalized in wall units rather than by the Kolmogorov length scale.

\subsection{NN-based SGS model trained with fDNS and fLES data: \textit{a priori} and \textit{a posteriori} tests}
As mentioned before, we train the NN with fLES data as well as fDNS data. The theoretical basis for filtering LES data can be attributed to the test filter introduced by \citet{Germano91}:
\begin{equation}
\widetilde{\phi}(\boldsymbol{x},t)=\int \widetilde{G}(\boldsymbol{r},\boldsymbol{x})\phi(\boldsymbol{x-r},t)d\boldsymbol{r},
\label{eq:lastone}
\end{equation}
where $\tilde G$ is a filter function and $\widetilde{\bar G} = \tilde G \bar G$. The test-filtered SGS stresses are written as 
\begin{equation}
T_{ij} = \widetilde{\overline{{ u}_i{ u}_j}}-\widetilde{\overline u}_i\widetilde{\overline u}_j,
\label{eq:twentyfour}
\end{equation}
\begin{equation}
{T}^r_{ij} = {T}_{ij}-\frac{1}{3}{T}_{kk}\delta_{ij},
\label{eq:twentyfive}
\end{equation}
and ${T}^r_{ij}$ is obtained by the following relations:
\begin{equation}
{T}^r_{ij} = \widetilde{\tau}^r_{ij}+{L}^r_{ij}, 
\label{eq:twentyeight}
\end{equation}
where
\begin{equation}
L_{ij} = \widetilde{\overline{ u}_{i} \overline{ u}_{j}}-\widetilde{\overline u}_i\widetilde{\overline u}_j,
\label{eq:twentysix}
\end{equation}
\begin{equation}
{L}^r_{ij} = {L}_{ij}-\frac{1}{3}{L}_{kk}\delta_{ij}.
\label{eq:twentyseven}
\end{equation}

\begin{table}
 \begin{center}
 \begin{tabular}{ll}
 SGS model & Training data \\[3pt]
 VGM128D & fDNS32/128, fDNS64/128 \\
 VGM256DD & fDNS64/256, fDNS128/256 \\
 VGM(128D+256D) & fDNS32/128, fDNS64/128, fDNS32/256 \\
 VGM(128D+256L) & fDNS32/128, fDNS64/128, fLES32/256 \\
 VGM(128D+1024L) & fDNS32/128, fDNS64/128, fLES32/256, fLES32/512, fLES32/1024\\
 VGM(128D+32768L) & fDNS32/128, fDNS64/128, fLES32/256, fLES32/512, fLES32/1024,\\
 & fLES32/2048, fLES32/4096, fLES32/8192, fLES32/16384, fLES32/32768 \\
 \end{tabular}
 \caption{VGMs and corresponding training data. Here, VGM128D and VGM256DD are the same as VGM(32+64)/128 and VGM(64+128)/256 discussed in \S \ref{sec:3_1}, respectively.}
 \label{tab:table6}
 \end{center}
\end{table}

The present VGM is updated through the accumulation of new datasets with the input of $\widetilde{\Delta}^2|\widetilde{\bar \alpha}|\widetilde{\bar \alpha}_{ij}$ and the output of ${T}^r_{ij}$. The test filter size is twice the LES grid size, i.e., $\widetilde{\Delta}=2\bar \Delta_{\rm LES}$. The filtered LES data, referred to as fLES32/256, are obtained by filtering LES64/256 data. The fLES data is seleted using the same normalization and undersampling techniques described in \S \ref{sec:3_1}. The fLES32/512, fLES32/1024, $\cdots$, fLES32/32768 data can be created by filtering LES64/512, LES64/1024, $\cdots$, LES64/32768, respectively, during the recursive process. Table \ref{tab:table6} summarizes the present VGMs considered and corresponding training data.

\begin{table}
  \begin{center}
    \begin{tabular}{lcccc}
      SGS model & $R(\tau_{11}^{r})$ & $R(\tau_{12}^{r})$ & $R(\epsilon_{SGS})$ & $\epsilon_{SGS}$ \\[3pt]
      fDNS32/256 & \protect\emptyline & \protect\emptyline & \protect\emptyline & 0.734 \\
      VGM(128D+256D) \qquad & 0.528 & 0.563 & 0.540 & 0.653 \\
      VGM(128D+256L) & 0.527 & 0.563 & 0.539 & 0.706 \\
      CSM32/256 & 0.196 & 0.221 & 0.489 & 1.397 \\
      DSM32/256 & 0.196 & 0.221 & 0.489 & 1.316 \\
      GM32/256 & 0.511 & 0.544 & 0.507 & 0.378 \\
    \end{tabular}
    \caption{Statistics from \textit{a priori} test at $Re_L=375.69$ with $\bar \Delta/\eta=16.76$: Pearson correlation coefficients $(R)$ of $\tau_{ij}^{r}$ and $\epsilon_{SGS}$, and magnitude of $\epsilon_{SGS}$.}
    \label{tab:table8}
  \end{center}
\end{table}

\begin{figure}
 \centerline{\includegraphics[width=1.0\textwidth]{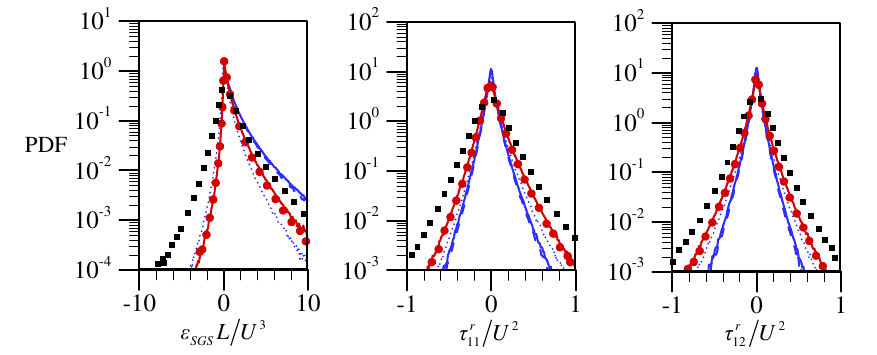}}
 \caption{Probability density functions  (\textit{a priori} test at $Re_L=375.69$ with $\bar \Delta/\eta=16.76$): (a) SGS dissipation; (b) SGS normal stress; (c) SGS shear stress. \protect\blasquare, fDNS32/256; \protect\redcircle, VGM(128D+256D); \protect\redline, VGM(128D+256L);  \protect\blueline, CSM32/256; \protect\bluedashedline, DSM32/256; \protect\bluedottedline, GM32/256.}
\label{fig:priori3}
\end{figure}

\begin{figure}\centerline{\includegraphics[width=1.0\textwidth]{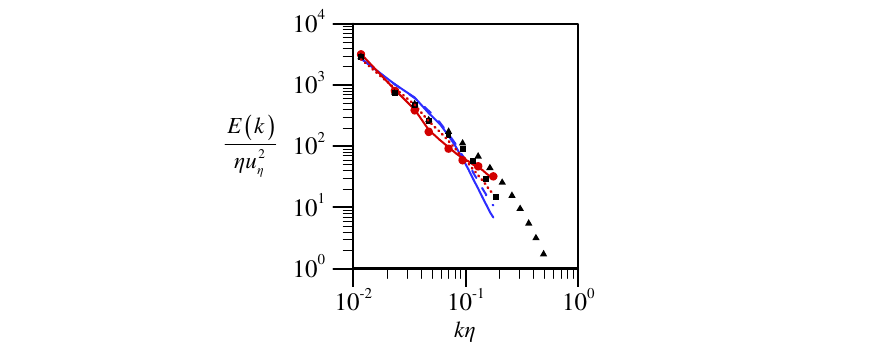}}
 \caption{Three-dimensional energy spectra (\textit{a posteriori} test at $Re_L=375.69$ with $\bar \Delta_{\rm LES}/\eta=16.76$): \protect\blatriangle, DNS256; \protect\blasquare, fDNS32/256; \protect\redcircle, VGM(128D+256D); \protect\redline, VGM(128D+256L); 
 \protect\reddottedline, VGM(128D+1024L);
 \protect\blueline, CSM32/256; \protect\bluedashedline, DSM32/256.}
\label{fig:Ek1d3}
\end{figure}

\begin{figure}\centerline{\includegraphics[width=1.0\textwidth]{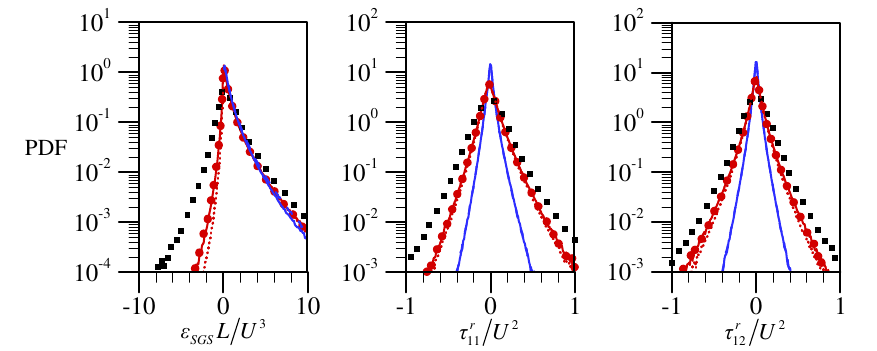}}
 \caption{Probability density functions  (\textit{a posteriori} test at $Re_L=375.69$ with $\bar \Delta_{\rm LES}/\eta=16.76$): (a) SGS dissipation; (b) SGS normal stress; (c) SGS shear stress. \protect\blasquare, fDNS32/256; \protect\redcircle, VGM(128D+256D); \protect\redline, VGM(128D+256L); 
 \protect\reddottedline, VGM(128D+1024L);
 \protect\blueline, CSM32/256; \protect\bluedashedline, DSM32/256.}
\label{fig:posteriori3}
\end{figure}

Let us conduct \textit{a priori} and \textit{a posteriori} tests for $Re_L=375.69$ with $\bar \Delta/\eta=16.76$ (twice that of LES64/256 and LES32/128). During the training process, VGM(128D+256D) employs fDNS data only (fDNS32/128, fDNS64/128 and fDNS32/256), but VGM(128D+256L) utilizes a combination of fDNS and fLES data (fDNS32/128, fDNS64/128 and fLES32/256). Table \ref{tab:table8} and figure \ref{fig:priori3} show the results of \textit{a priori} tests. As shown, the performances of VGMs are similar to that of GM, and are better than those of CSM and DSM. A notable observation is that the results of VGM(128D+256L) are nearly identical to those obtained from VGM(128D+256D), implying that fLES data can replace fDNS data in constructing NN-based SGS models.
Figures \ref{fig:Ek1d3} and \ref{fig:posteriori3} show the results of energy spectra and PDFs from \textit{a posteriori} tests, respectively, where the result from GM32/256 is not shown because the simulation diverged. Again, VGM(128D+256D) and VGM(128D+256L) perform well among the SGS models considered, but show underpredictions at intermediate wavenumbers. Thus, we perform additional LES with VGM(128D+1024L) whose trained filter sizes are $4.19 \le \bar \Delta_{\rm fDNS}/\eta$ and $\bar \Delta_{\rm fLES}/\eta \le 67.02$ (note that the filter sizes for VGM(128D+256L) are $\bar \Delta_{\rm fDNS}/\eta$ and $\bar \Delta_{\rm fLES}/\eta = 4.19 \sim 16.76$).
The results with VGM(128D+1024L) are in excellent agreements with fDNS32/256, suggesting that one should expect successful LES when $\bar \Delta_{\rm LES} / \eta$ is within the range of trained filter sizes (see \S \ref{sec:4.2} for further discussion).
It is also notable that the present VGMs do not show the inconsistency between \textit{a priori} and \textit{a posteriori} tests observed from the traditional SGS models \citep{Park05}.

\section{Applications to FHIT at high Reynolds numbers and DHIT}
\label{sec:4}
To assess the applicability of the present VGMs (table \ref{tab:table6}), we conduct LESs of FHIT at high Reynolds numbers and of DHIT. 

\subsection{Forced homogeneous isotropic turbulence at high Reynolds numbers}

\begin{figure}
  \centerline{\includegraphics[width=0.9\textwidth]{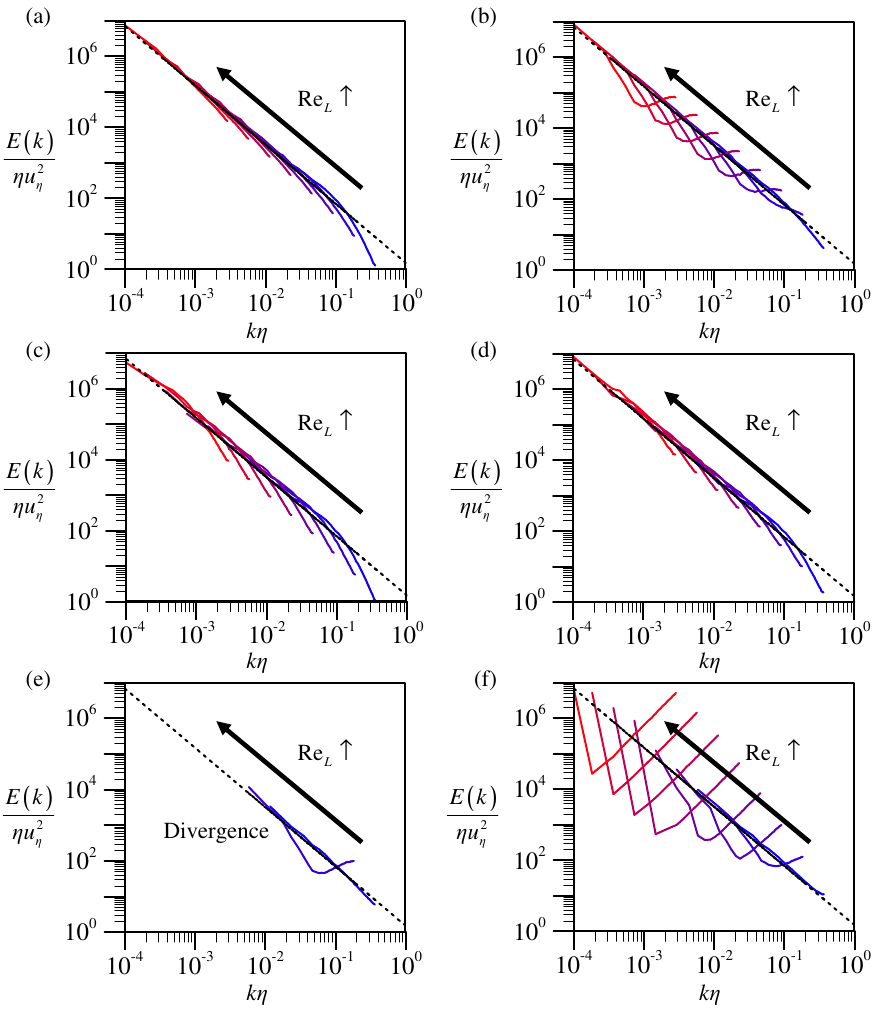}}
  \caption{Three-dimensional energy spectra at $Re_L = 375.69 - 242347.33$ from various SGS models: (a) VGM(128D+32768L); (b) VGM128D; (c) CSM; (d) DSM; (e) GM; (f) no SGS model. The black dashed line (\protect\dashed) denotes the Kolmogorov energy spectrum, $E(k)=\frac{3}{2}{\epsilon}^{2/3} {k}^{-5/3}$.}
  \label{fig:Recursive}
\end{figure}

LESs of FHIT with the number of grid points of $64^3$ (LES64/256, LES64/512, LES64/1024, LES64/2048, LES64/4096, LES64/8192, LES64/16384, and LES64/32768) are conducted at eight different Reynolds numbers ($Re_{L} = 375.69 - 242347.33$) using five different SGS models (VGM128D, VGM(128D+32768L), CSM, DSM, and GM). 
The computational time step is set at $\Delta t U/L=0.0025$, at which the Courant–Friedrichs–Lewy (CFL) numbers are below 0.3. Figure \ref{fig:Recursive} shows the three-dimensional energy spectra at eight Reynolds numbers from various SGS models, together with the Kolmogorov energy spectrum. First, LESs with GM reveal instability at high Reynolds numbers. 
LESs without SGS model exhibit non-physical energy pile up at high wavenumbers (figure \ref{fig:Recursive}(f)) due to insufficient dissipation. A similar pile up of energy at high wavenumbers is also observed in LESs with VGM128D (figure \ref{fig:Recursive}(b)).
The results from VGM128D indicate an inherent limitation of non-recursive NN-based SGS model, when the LES grid size is bigger than the filter sizes of fDNS employed for training the NN.
On the other hand, VGM(128D+32768L) suitably calculates the energy spectra that are quite similar to those of CSM and DSM (figures \ref{fig:Recursive}(c) and (d)). The predicted energy spectra from VGM, CSM and DSM follow the Kolmogorov energy spectrum at low and intermediate wavenumbers. 
 
\subsection{Decaying homogeneous isotropic turbulence}\label{sec:4.2}
\begin{figure}
 \centerline{\includegraphics[width=0.9\textwidth]{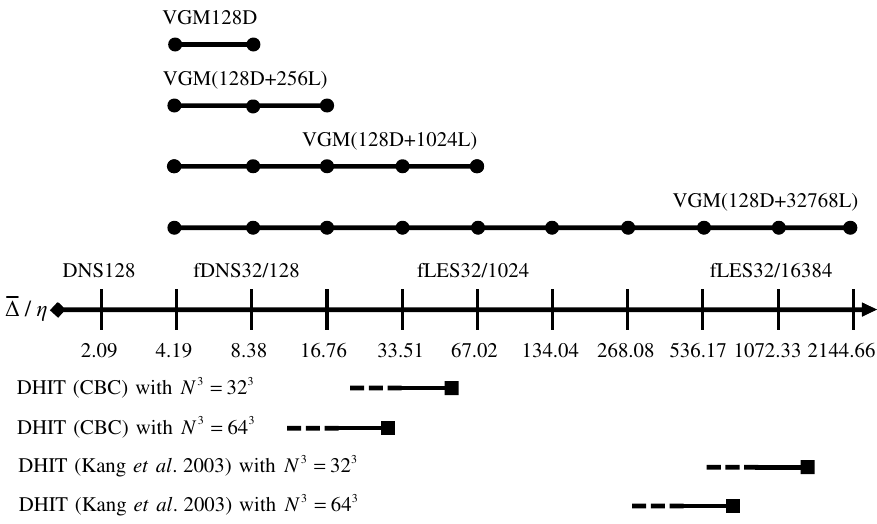}}
 \caption{Illustration of the filter-size ranges of fDNS and fLES used to generate each VGM (FHIT), and the grid-size ranges required for the simulations of DHITs \citep{Comte71,Kang03}. On the top of this figure (FHIT), each solid circle (\protect\blackcircle) denotes the filter size of fDNS or fLES used for generating training data. For each VGM, the line connecting the first and last solid circles indicates the range of grid sizes for successful LES. On the bottom of this figure (DHIT), each solid square (\protect\blasquare) indicates the grid size normalized by the initial Kolmogorov length scale. As time goes by, the Kolmogorov length scale increases and thus $\bar{\Delta}_{\rm LES}/\eta$ decreases denoted as a thick dashed line.}
\label{fig:gridsize}
\end{figure}

The present approach is based on a hypothesis that the NN-based SGS models in LES successfully predict the turbulence statistics when the LES grid size normalized by the Kolmogorov length scale is within the range of filter sizes used for training (see also \cite{Park21}). In figure \ref{fig:gridsize}, we indicate the filter sizes (normalized by the Kolmogorov length scale) used for LESs of FHIT with VGMs and the normalized grid sizes used for promising LESs of two DHITs, respectively. This figure should be understood as follows: for DHIT of \cite{Comte71}, VGM(128D+1024L) or VGM(128D+32768L) can be used for successful LES, but not VGM128D and VGM(128D+256L); for DHIT of \cite{Kang03}, only VGM(128D+32768L) can be used for successful LES. To examine this hypothesis, LESs are performed for two DHITs of \cite{Comte71} and \cite{Kang03} with the present VGMs trained with FHIT data. This task should be a notable challenge due to transient nature of DHIT as opposed to FHIT.

The first DHIT to simulate is the experiment by \cite{Comte71} (called CBC hereafter). In CBC, turbulence data were experimentally generated through grid turbulence using a mesh size of $M=5.08$ cm and free-stream velocity of $U_0=10$ m/s. The Reynolds number based on the Taylor integral length scale was $Re_{\lambda}=71.6$ at $tU_0/M=42$. For LES, flow variables are non-dimensionalized by $L=11M /(2\pi)$ and $U=\sqrt{3/2} ~u_{rms} \vert_{tU_0/M=42}$ \citep{Lee10}. The numbers of grid points tested are $N^3=32^3$ and $64^3$, respectively. A divergence-free initial field is obtained using logistic polynomial approximation method \citep{knight98} and rescaling method \citep{Kang03}. LESs are performed 30 times by varying initial fields to obtain ensemble averages. The computational time step is fixed to be ${\Delta} t U / L = 0.01$. As turbulence decays in time, the Kolmogorov length scale $\eta$ increases in time, and thus $\bar \Delta_{\rm LES}/\eta = 60.2$ and 26.5 ($N^3 = 32^3$), and $30.1$ and $13.2$ ($N^3 = 64^3$) at $t U_0 / M=42$ and $171$, respectively. 
Each LES starts from large $\bar{\Delta}_{LES}/\eta$ (denoted as a solid square in figure \ref{fig:gridsize}) and moves to smaller $\bar{\Delta}_{LES}/\eta$ due to increased $\eta$ in time (denoted as a thick dashed line). 

\begin{figure}
 \centerline{\includegraphics[width=0.9\textwidth]{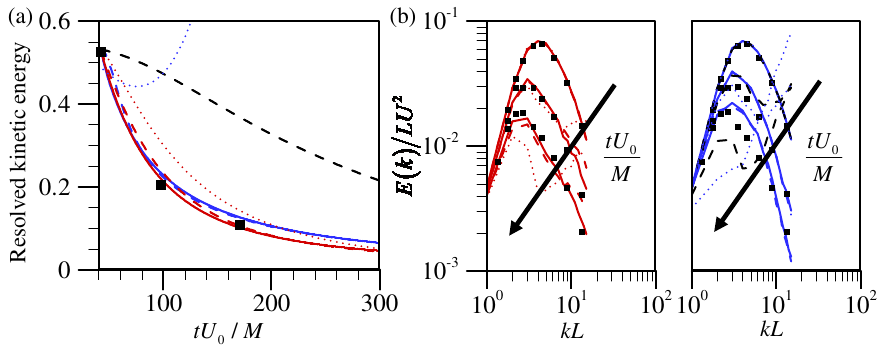}}
 \caption{LES of CBC DHIT with $N^3 = 32^3$: (a) resolved turbulent kinetic energy; (b) $E(k)$ at $tU_{0}/M=42, 98$ and $171$ from VGM128D, VGM(128D+1024L), VGM(128D+32768L), CSM, DSM, GM and no SGS model. \protect\blasquare, Filtered CBC data; \protect\reddottedline, VGM128D;  \protect\reddashedline, VGM(128D+1024L);  \protect\redline, VGM(128D+32768L);  \protect\blueline, CSM; \protect\bluedashedline, DSM; \protect\bluedottedline, GM; \protect\dashedline, no SGS model.}
\label{fig:CBC32}
\end{figure}

\begin{figure}
 \centerline{\includegraphics[width=0.9\textwidth]{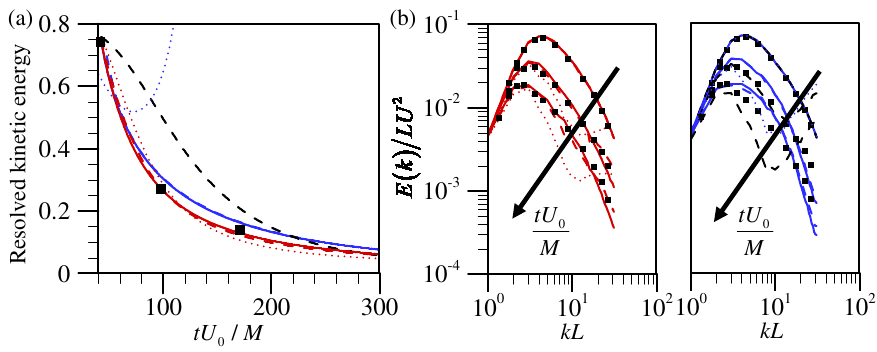}}
 \caption{LES of CBC DHIT with $N^3 = 64^3$: (a) resolved turbulent kinetic energy; (b) $E(k)$ at $tU_{0}/M=42, 98$ and $171$ from VGM128D, VGM(128D+1024L), VGM(128D+32768L), CSM, DSM, GM and no SGS model.  \protect\blasquare, Filtered CBC data; \protect\reddottedline, VGM128D; \protect\reddashedline, VGM(128D+1024L);  \protect\redline, VGM(128D+32768L);  \protect\blueline, CSM; \protect\bluedashedline, DSM; \protect\bluedottedline, GM; \protect\dashedline, no-SGS model.}
\label{fig:CBC64}
\end{figure}

The LES results on CBC DHIT with $N^3 = 32^{3}$ (starting from $\bar \Delta_{\rm LES} / \eta = 60.2$) are given in figure \ref{fig:CBC32}. Without SGS model, the turbulent kinetic energy decays very slowly and energy pile up occurs at high wavenumbers. With GM, resolved kinetic energy first decreases and increases later in time, and simulation finally diverges, as shown in \citet{Vreman96}. DSM and CSM are relatively good at predicting turbulence decay and energy spectra, but they overestimate the energy spectra at intermediate wavenumbers within the inertial range. 
On the other hand, VGM(128D+32768L) performs best in predicting energy spectra, while VGM(128D+1024L) also performs better than other types of SGS models but shows energy pile up at high wavenumbers, possibly because $\bar \Delta_{\rm LES}/ \eta ~(\le 60.2)$ is only slightly smaller than $\bar \Delta_{\rm fLES} / \eta=67.02$ (fLES32/1024). Note also that the prediction with VGM128D ($\bar \Delta_{\rm LES} / \eta \gg \bar \Delta_{\rm fDNS} / \eta=8.38$) is not good at all and worse than CSM and DSM, validating our conjecture on the choice of VGM for successful LES (figure \ref{fig:gridsize}). With $N^3 = 64^3$ (starting from $\bar \Delta_{\rm LES} / \eta = 30.1$), VGM(128D+1024L) also performs very good as much as VGM(128D+32768L) does (figure \ref{fig:CBC64}), because $\bar \Delta_{\rm LES} / \eta ~(\le 30.1)$ is sufficiently smaller than $\bar \Delta_{\rm fLES} / \eta=67.02$, while other SGS models still show some deviations from experimental data.

The second DHIT to simulate is the experimental one by \citet{Kang03}. The Reynolds number is much higher ($Re_{\lambda}=716$ at $tU_{0}/M=20$) than that of CBC DHIT, where $U_{0}=11.2$ m/s and $M=0.152$ m. We examine five SGS models (VGM(128D+32768L), VGM128D, CSM, DSM, and GM), and LESs are performed 30 times by varying initial fields to obtain ensemble averages. 
The flow variables are non-dimensionalized by $L  
= 33.68M/(2\pi)$ and $U=\sqrt{3/2} ~u_{rms} \vert_{tU_0/M=20}$. The numbers of grid points tested are $N^3=32^3$ and $64^3$, respectively. 
The flows are initiated using the energy spectrum and rescaling method as done in \citet{Kang03}. For $N^3 = 32^3$, $\bar \Delta_{\rm LES}/\eta=1454.5$ and 888.9 at $tU_{0}/M=20$ and $48$, respectively. The trained grid sizes of VGM(128D+32768L) ($4.19 \le \bar \Delta_{\rm fDNS} / \eta$ and $\bar \Delta_{\rm fLES} / \eta \le 2144.66$) include this range of $\bar \Delta_{\rm LES}/\eta$, whereas those of VGM128D ($4.19 \le \bar \Delta_{\rm fDNS} / \eta \le 8.38$) do not (see figure \ref{fig:gridsize}), suggesting that VGM(128D+32768L) should properly predict turbulence decay and variation of the energy spectra of DHIT by \cite{Kang03}.

\begin{figure}
 \centerline{\includegraphics[width=0.9\textwidth]{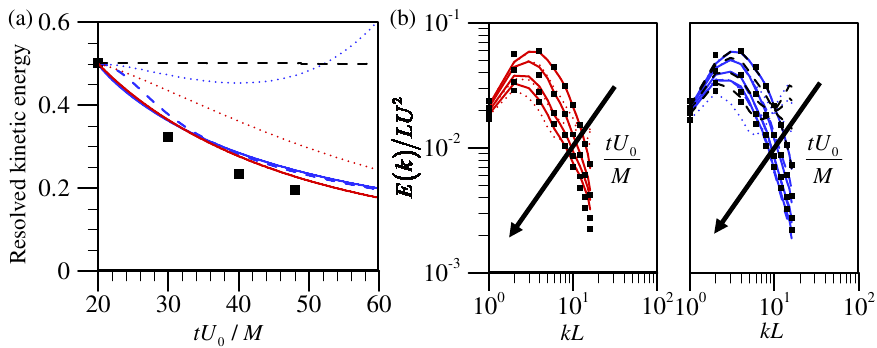}}
 \caption{LES of DHIT by \cite{Kang03} with $N^3 = 32^3$: (a) resolved turbulent kinetic energy; (b) $E(k)$ at $tU_{0}/M= 20, 30, 40$ and $48$ from VGM128D, VGM(128D+32768L), CSM, DSM, GM and no SGS model. \protect\blasquare, Filtered experimental data; \protect\reddottedline, VGM128D; \protect\redline, VGM(128D+32768L); \protect\blueline, CSM; \protect\bluedashedline, DSM; \protect\bluedottedline, GM; \protect\dashedline, no-SGS model.}
\label{fig:Kang32}
\end{figure}

\begin{figure}
 \centerline{\includegraphics[width=0.9\textwidth]{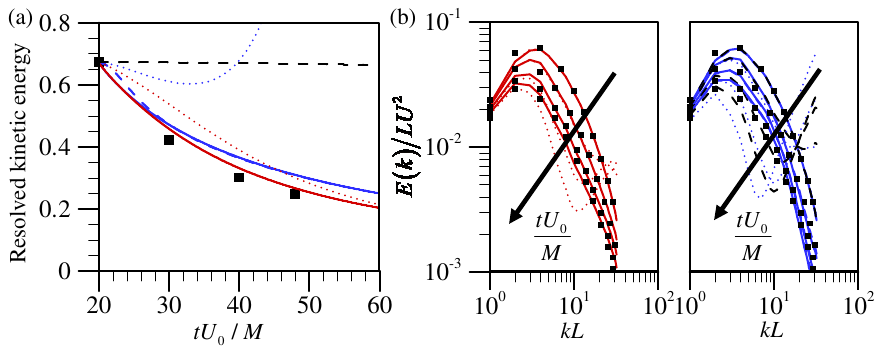}}
 \caption{LES of DHIT \citep{Kang03} with $N^3 = 64^3$: (a) resolved turbulent kinetic energy; (b) $E(k)$ at $tU_{0}/M=20, 30, 40$ and $48$ from CSM, DSM, GM and no SGS model;  (c) from VGM128D and VGM(128D+32768L). \protect\blasquare, Filtered experimental data; \protect\reddottedline, VGM128D;  \protect\redline, VGM(128D+32768L);  \protect\blueline, CSM; \protect\bluedashedline, DSM; \protect\bluedottedline, GM; \protect\dashedline, no-SGS model.}
\label{fig:Kang64}
\end{figure}

The LES results on DHIT by \cite{Kang03} with $N^3 = 32^3$ are shown in figure \ref{fig:Kang32}. While CSM and DSM overpredict the energy spectra even at intermediate wavenumbers and thus overestimate turbulent kinetic energy, VGM(128D+32768L) predicts the resolved kinetic energy and energy spectra most accurately. However, VGM(128D+32768L) shows slight overprediction of energy at high wavenumbers, causing higher resolved kinetic energy than the experimental one. The prediction behaviours of SGS models considered are very similar to what we observed from CBC DHIT. By increasing the number of grid points ($N^3=64^3$), the predictions by VGM(128D+32768L) become much better (figure \ref{fig:Kang64}). 
It is interesting to see that the level of kinetic energy without SGS model is almost constant in time due to the energy pile up at high wavenumbers. This is because the LES grid size ($\Delta_{\rm LES}/\eta \approx 10^3$) is so large that the correspoding eddies do not properly dissipate energy at these length scales, and thus total kinetic energy is nearly conserved without SGS model.

\section{Conclusions}
\label{sec:5}
In the present study, we developed an NN-based SGS model (velocity gradient model; VGM) using a dual NN architecture (the output of one NN is the SGS normal stresses and that of the other is the SGS shear stresses, respectively) with the input of $\bar{\Delta}^2|\bar{\alpha}|\bar \alpha_{ij}$, where $\alpha_{ij}$ is the velocity gradient tensor.
By eliminating bias and employing the leaky ReLU function as an activation function within the NNs, the present NN retains its nonlinearity but satisfies $\boldsymbol{{\mathrm{NN}}} (c\boldsymbol{A}) = c \boldsymbol{{\mathrm{NN}}} (\boldsymbol{A})$ for a positive scalar $c$. This important property of NN allows us to apply the NN to a flow different from the trained one, because the parameters used for non-dimensionalization can be included in the scalar $c$. 
To generate training data, we adopted a cut-Gaussian filter rather than the Gaussian or spectral cut-off filter, considering its application to LES with coarser grids and realizability condition, respectively. 
We also designed a recursive procedure which consisted of the following steps: (1) conducting DNS; (2) training an NN with fDNS data; (3) conducting LES at a higher Reynolds number; (4) training the NN with augmented data including fLES data; (5) going to (3) for higher Reynolds number flows. For the present FHIT, the grid and filter sizes were normalized by the Kolmogorov length scale, and these normalized sizes became double at every recursive procedure. 
The NN trained through this recursive procedure contained a wide range of filter sizes normalized by the Kolmogorov length scale such that the LES grid size required could be located within this range of filter sizes.
Testing VGMs on FHIT showed that fLES data can be a practical alternative to fDNS data for training an NN, and one can avoid using costly DNS to extract training data.

We conducted LESs of forced and decaying homogeneous isotropic turbulence (FHIT and DHIT) with VGMs and traditional SGS models, respectively. In FHIT, the same number of grid points was used for different Reynolds numbers. Among the SGS models considered, VGM constructed through the recursive process performed very well for FHIT. In contrast, VGM trained only with fDNS data at a low Reynolds number showed energy pile up at high wavenumbers. We considered two different DHIT, one by \citet{Comte71} and the other at much higher Reynolds number by \cite{Kang03}. In both DHIT, the recursive VGM predicted the decay of resolved turbulent kinetic energy and its energy spectra in time most accurately among the SGS models considered, while other SGS models provided excessive energy spectra at intermediate or high wavenumbers.

In the present study, we applied the VGM to FHIT and DHIT, but this procedure may not be applicable to laminar and inhomogeneous turbulent flows. Hence, the next step is to extend the present approach to those flows by employing dynamic approach used in DSM or training NN with more flows having different topology.  This work is being carried out in our group \citep{Kim23, Cho23}.

\bigskip
\noindent
Declaration of Interests. The authors report no conflict of interest.

\bigskip
\noindent
\textbf{Acknowledgements}
This work is supported by the National Research Foundation through the Ministry of Science and ICT (no. 2022R1A2B5B0200158612 and 2021R1A4A1032023). The computing resources are provided by the KISTI Super Computing Center (no. KSC-2023-CRE-0197).

\bibliographystyle{jfm}
\bibliography{references}

\end{document}